\documentclass[12pt,aps,prapplied,reprint, showpacs,showkeys, superscriptaddress]{revtex4-1}
\usepackage{newcent}

\usepackage[T1]{fontenc}
\usepackage[latin9]{inputenc}
\usepackage{textcomp}
\usepackage{amsmath}
\usepackage{commath}
\usepackage{csquotes}
\usepackage{graphicx, subfigure}
\usepackage{amssymb}
\usepackage{amsbsy}
\usepackage{textcomp}
\usepackage{eucal}
\usepackage{siunitx}
\usepackage{textcomp}
\usepackage{pslatex}
\usepackage{relsize}
\usepackage{bm}
\usepackage{color}
\usepackage{hyperref}
\hypersetup{colorlinks=true,linkcolor=blue,citecolor=blue,urlcolor=blue}
\usepackage[section]{placeins} 

\usepackage[normalem]{ulem}

\begin{document}

\title{Effective macrospin model for $Co_{x}Fe_{3-x}O_{4}$ nanoparticles: decreasing the anisotropy by Co-doping?}

\author{David Serantes}
\email[E-mail: ]{david.serantes@usc.es}
\address{Instituto de Investigaci\'ons Tecnol\'oxicas and Applied Physics Department, Universidade de Santiago de Compostela, 15782 Santiago de Compostela, Spain}
\author{Daniel Fa\'ilde}
\address{Instituto de Investigaci\'ons Tecnol\'oxicas and Applied Physics Department, Universidade de Santiago de Compostela, 15782 Santiago de Compostela, Spain}
\author{Daniel Baldomir}
\address{Instituto de Investigaci\'ons Tecnol\'oxicas and Applied Physics Department, Universidade de Santiago de Compostela, 15782 Santiago de Compostela, Spain}

\author{Beatriz Pelaz}
\address{Centro Singular de Investigaci\'on en Qu\'imica 
Biol\'oxica e Materiais Moleculares (CiQUS), Departamento de F\'isica de Part\'iculas, Universidade de Santiago de Compostela, 15782 Santiago de Compostela, Spain} 

\author{Pablo del Pino}
\address{Centro Singular de Investigaci\'on en Qu\'imica 
Biol\'oxica e Materiais Moleculares (CiQUS), Departamento de F\'isica de Part\'iculas, Universidade de Santiago de Compostela, 15782 Santiago de Compostela, Spain} 

\author{Roy W. Chantrell}
\address{Department of Physics, University of York,York,
YO10 5DD,
United Kingdom}

\begin{abstract}
$Co$-doping of $Fe_{3}O_{4}$ magnetic nanoparticles is an effective way to tailor their magnetic properties. When considering the two extreme cases of the $Co_{x}Fe_{3-x}O_{4}$ series, i.e. the $x=0$ and $x=1$ values, one finds that the system evolves from a negative cubic-anisotropy energy constant, $K_{C}^{-}<0$, to a positive one, $K_{C}^{+}>0$. Thus, what happens for intermediate $x$-compositions? In this work we present a very simple phenomenological model for the anisotropy, under the \textit{macrospin} approximation, in which the resultant anisotropy is just directly proportional to the amount of $Co$. First, we perform a detailed analysis on a rather ideal system in which the extreme values have the same magnitude (i.e. $|K_{C}^{-}|=|K_{C}^{+}|$) and then we focus on the real $Co_{x}Fe_{3-x}O_{4}$ system, for which $|K_{C}^{+}|\sim 18|K_{C}^{-}|$. Remarkably, the approach  reproduces rather well the experimental values of the heating performance of $Co_{x}Fe_{3-x}O_{4}$ nanoparticles, suggesting that our simple approach may in fact be a good representation of the real situation. This gives rise to an intriguing related possibility arises: a $Co$-doping composition should exist for which the effective anisotropy tends to zero, estimated here as 0.05.
\end{abstract}

\maketitle

\section{Introduction}
\begin{figure}[!ht]
\includegraphics[width = 0.8\textwidth]{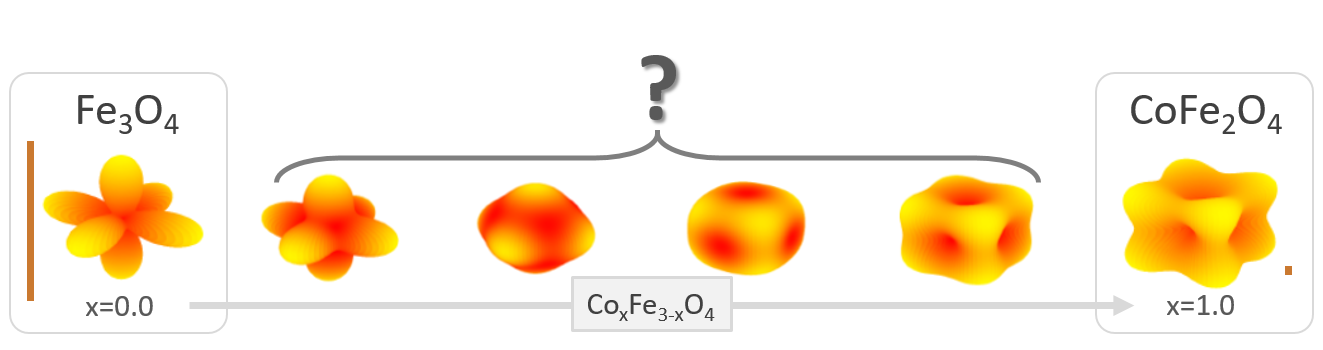}
\caption{Schematic illustration of the evolution of the effective anisotropy landscape of $Co_{x}Fe_{3-x}O_{4}$ nanoparticles as a function of the $Co$-content $x$. The bars for $x=0$ and $x=1$ represent the relative magnitudes of the anisotropy in each case. }
\label{fig:figure_1}
\end{figure}

To have particles with similar magnetic moment, shape and size, but different anisotropy, a possible experimental approach is to use $Fe_{3}O_{4}$ magnetite nanoparticles as a base system, and to dope them with e.g. $Mn$ or $Co$. In this way, a large variation in the anisotropy constant $K$ is expected while maintaining similar values of the other characteristic parameters. However, from the theoretical point of view, a puzzling question arises regarding the magnetic anisotropy of the system. Consider for example the $Co_{x}Fe_{3-x}O_{4}$ series as a function of $x$, which has cubic \textit{negative} anisotropy constant $K_{C}^{-}<0$ on the $x=0$ extreme of the series (corresponding to $Fe_{3}O_{4}$), but cubic \textit{positive} anisotropy $K_{C}^{+}>0$ on the other extreme $x=1$ (i.e. for $CoFe_{2}O_{4}$). The question is, what happens for intermediate compositions? What would be the \textit{effective} anisotropy of the Co-doped $Fe_{3}O_{4}$ nanoparticles? The magnetic anisotropy is a key magnetic parameter, particularly for nano-scaled materials at the single-domain range, as it determines the dynamical behaviour\cite{Livesey2018, Serantes2010}, efficiency of the magnetic torque \cite{Chantrell1983,Serantes2018}, or stability of ordered assemblies\cite{Serantes2009,Arora2019}. Thus, controlling the anisotropy is essential for a broad range of applications as diverse as magnetic recording\cite{Richter2012, Aas2013}, hyperthermia \cite{Conde-Leboran2015,Ruta2015}, or magnetic refrigeration\cite{vonRanke2008, Serantes2012}. In the following we will develop in this context the \textit{effective} anisotropy from a single-particle (\textit{macrospin} approximation) approach. The article is organized as follows: in Section \ref{the_idea} we introduce our simple idea to treat this problem; essentially, to approximate the effective anisotropy as dependent on the amount of $Co$-doping (see Figure \ref{fig:figure_1}). Then, in Section \ref{the_model} we describe the computational model used to study the problem. The results are reported and analysed in Section \ref{results}, which contains 2 subsections: firstly, in \ref{ideal_case} we analyse the role of the symmetry of the anisotropy, for the ideal case in which $|K_{C}^{-}|=|K_{C}^{+}|$; then, in \ref{real_case} the real $Co_{x}Fe_{3-x}O_{4}$ case is presented. Finally, Section \ref{conclusions} presents some conclusions and a summary of the work.

\onecolumngrid

\twocolumngrid

\section{The phenomenological idea}\label{the_idea}

For simplicity, we can envisage a nanoparticle small enough  that the magnetic behaviour of the atomic moments is dominated by exchange. In this case, magnetic properties are appropriately described by the \textit{macrospin} approximation -i.e. the particle behaves as a large \textit{supermoment}. For such a nanoparticle, what would be the effect on the net anisotropy of having some $Fe$ sites replaced with $Co$? Here we assume firstly  that $Co^{2+}$ cations displace octahedral $Fe^{2+}$ sites\cite{Slonczewski1958}. Further we assume that, since the anisotropy is a local property, to a first approximation we postulate the coexistence of both types of anisotropies. Then it follows that the net anisotropy is the volume weighted average of the anisotropies of the sites with $Fe^{2+}$ and $Co^{2+}$ cations.

To test this idea we developed a simple theoretical approach in which a positive cubic-anisotropy contribution directly proportional to the $Co$-doping fraction appears onto the cubic-negative anisotropy of the base $Fe_{3}O_{4}$ material. A systematic analysis and characterisation of the resulting magnetic properties of the model is carried out in general terms, and then applied to a particular case: the heating properties of $Co$-doped $Fe_{3}O_{4}$ magnetite nanoparticles as reported experimentally by Fantechi \textit{et al.}\cite{fantechi2015}. Finally, the intriguing possibility that a small $Co$-doping can lead to a decrease of the particle effective anisotropy is analyzed.

Assuming that the particles are sufficiantly small that the inner magnetic moments behave coherently, it seems reasonable to consider that the contribution of every magnetic moment will have a direct influence on the \textit{global} behaviour of the magnetic \textit{supermoment}. Thus, knowing that doping $Fe_{3}O_{4}$ with $Co$ adds a positive cubic anisotropy contribution\cite{Marcano2018}, our phenomenological approach gives that the effective particle anisotropy is, in first approximation, proportional to the volume weighted $Co$ and $Fe$ fractions as
\begin{equation}\label{combined_cubic_energies}
    K_{C}=K_{C}^{Fe_{3}O_{4}}\cdot\left({1-x}\right)+K_{C}^{CoFe_{2}O_{4}}\cdot{x} ,
\end{equation} with the nominal bulk anisotropies of $Fe_{3}O_{4}$ and $CoFe_{2}O_{4}$ given by $K_{C}^{Fe_{3}O_{4}}=-1.1\cdot{10^4} J/m^3$ and $K_{C}^{CoFe_{2}O_{4}}=2.0\cdot{10^5} J/m^3$, respectively. It is important to emphasize here that while our approach might appear similar to other works tailoring the effective anisotropy by combining soft and hard magnetic materials (e.g. in core/shell geometry, considering the contributions from the different layers to be additive \cite{Zhang2015, Carriao2016}), the  essence of the current work is however completely different. What is crucial here is introducing the mixed anisotropies of the $Co$ and $Fe$ at the atomic level which allows  fine-tuning of the anisotropy magnitude and easy-axis orientation, to the best or our knowledge not considered in other approaches. The condition of additivity of the properties  arises naturally as a consequence of the uniform magnetisation imposed by the nanoparticle dimensions.

As mentioned in the Introduction section, the simulated particle properties will be based on $Fe_{3}O_{4}$ with progressive $Co$ doping. For the sake of simplicity we have assumed that the saturation magnetisation does not vary significantly with $Co$ content\cite{Byrne2013}, considering it to have a constant value of $M_{S}=480 emu/cm^3$ for all cases. This reasonable assumption allows us to specifically concentrate on the role of the anisotropy. To ensure fully coherent magnetisation reversal we have focused on relatively small spherical nanoparticles of diameter $d=8.5 nm$; the same values as reported by Fantechi \textit{et al.}\cite{fantechi2015}. 

\section{Computational model}\label{the_model}

Given the small size of the particles, well below the single-domain threshold, we model them as large macrospins with coherent rotation of the inner magnetic moments, as often assumed in theoretical studies of very small magnetic nanoparticles\cite{Serantes2012,Ruta2015,Livesey2018}. Thus, the free energy of each particle \textit{i} is governed by the Zeeman and anisotropy energies as
\begin{equation}\label{cubic_energy}
    E_{i}=\vec{\mu_{i}}\cdot{\vec{H}}+K_{C}\left({\alpha^2\beta^2+\beta^2\gamma^2+\gamma^2\alpha^2}\right),
\end{equation} 
with $\abs{\vec{\mu_{i}}}=M_{S}V_{i}$, where $M_{S}$ is the saturation magnetisation and $V_{i}$ is the particle volume; $\vec{H}$ is the applied magnetic field; $K_{C}$ is the cubic anisotropy constant and $\alpha,\beta,\gamma$ are the cosine directors. The system is considered monodisperse in size, so that the only difference between particles is the orientation of the anisotropy axes and magnetic moment.

The dynamical magnetic response of the particles to the applied magnetic field is described by the Landau-Lifshitz-Gilbert equation using the OOMMF software package\cite{oommf}. In general the results will consist of simulating $M(H)$ hysteresis loops, both in the quasistatic limit (considering major loops) at zero temperature ($T=0$); and under rapidly time-varying AC fields (at room temperature) as  used in magnetic hyperthermia experiments. The quasistatic simulations were carried out by means of energy minimization; for the dynamical calculations we   used the Oxs Extension Module \textit{thetaevolve}\cite{oommf-thetaevolve}, to account for thermal effects. It must be noted that to avoid considerations regarding dispersion of particle parameters (size\cite{Munoz-Menendez2015}, anisotropy\cite{Munoz-Menendez2017}, etc), we have for the sake of simplicity assumed in all calculations perfectly monodisperse systems. In this regard, given the relatively small particle sizes of the target experiment\cite{fantechi2015}, of about $8.5 nm$ in diameter - thus with a likely significantly bigger heating performance for the larger particles\cite{Munoz-Menendez2015, Munoz-Menendez2016}, we used the following characteristic size. The simulated size $d_{simul}$ was chosen as the average mean value of each experimental case, $<d_{exp}>$, plus $1.5$ times the corresponding standard deviation ($\sigma$) of each sample, i.e. $d_{simul}=<d_{exp}>+1.5\sigma_{exp}$ to account for the higher weighting of larger particles.

\section{Results and discussion}\label{results}

To study the properties of the simple approach described by \eqref{combined_cubic_energies}, we shall simulate the magnetic properties of nanoparticles of different $Co_{x}Fe_{3-x}O_{4}$ compositions as a function of the $x$-fraction of $Co$ content. However, based on the large difference between the respective anisotropy constants of $Fe_{3}O_{4}$ and $CoFe_{2}O_{4}$ (thus with an expected much higher influence of the $K_{C}>0$ contribution), for illustrative purposes we will consider at first the ideal case of the anisotropy evolving between $K_{C}^{-}$ (x=0.0) and $K_{C}^{+}$ (x=1.0), with $|K_{C}^{-}|=|K_{C}^{+}|$. Later, the realistic case of the transition between $Fe_{3}O_{4}$ and $CoFe_{2}O_{4}$ alloys will be considered.

Thus, for now we will just consider the effective anisotropy $K_{C}^{eff}$ of the ideal situation described by
\begin{equation}\label{combined_NEG_POS-ideal}
    K_{C}^{eff}=K_{C}^{-}\cdot\left({1-x}\right)+K_{C}^{+}\cdot{x},
\end{equation}
where $K_{C}^{-}=-1.1\cdot{10^4} J/m^3$ and $K_{C}^{+}=1.1\cdot{10^4} J/m^3$. By doing so the colors depicted in Figure \ref{fig:figure_1} correspond to the same energy scale, thus making easy to discern the effect of the combined anisotropies. It is important to keep in mind that the different energy geometries are shown for illustrative purposes; otherwise the same geometry with inter-exchanged color scheme would be enough to resemble both positive and negative anisotropy energies.

\subsection{Ideal case: from $K_{C}^{-}$ to $K_{C}^{+}$, with $\lvert{K_{C}^{-}}\rvert=\lvert{K_{C}^{+}}\rvert$}\label{ideal_case}

\begin{figure}[ht]
\includegraphics[width = 0.6\columnwidth]{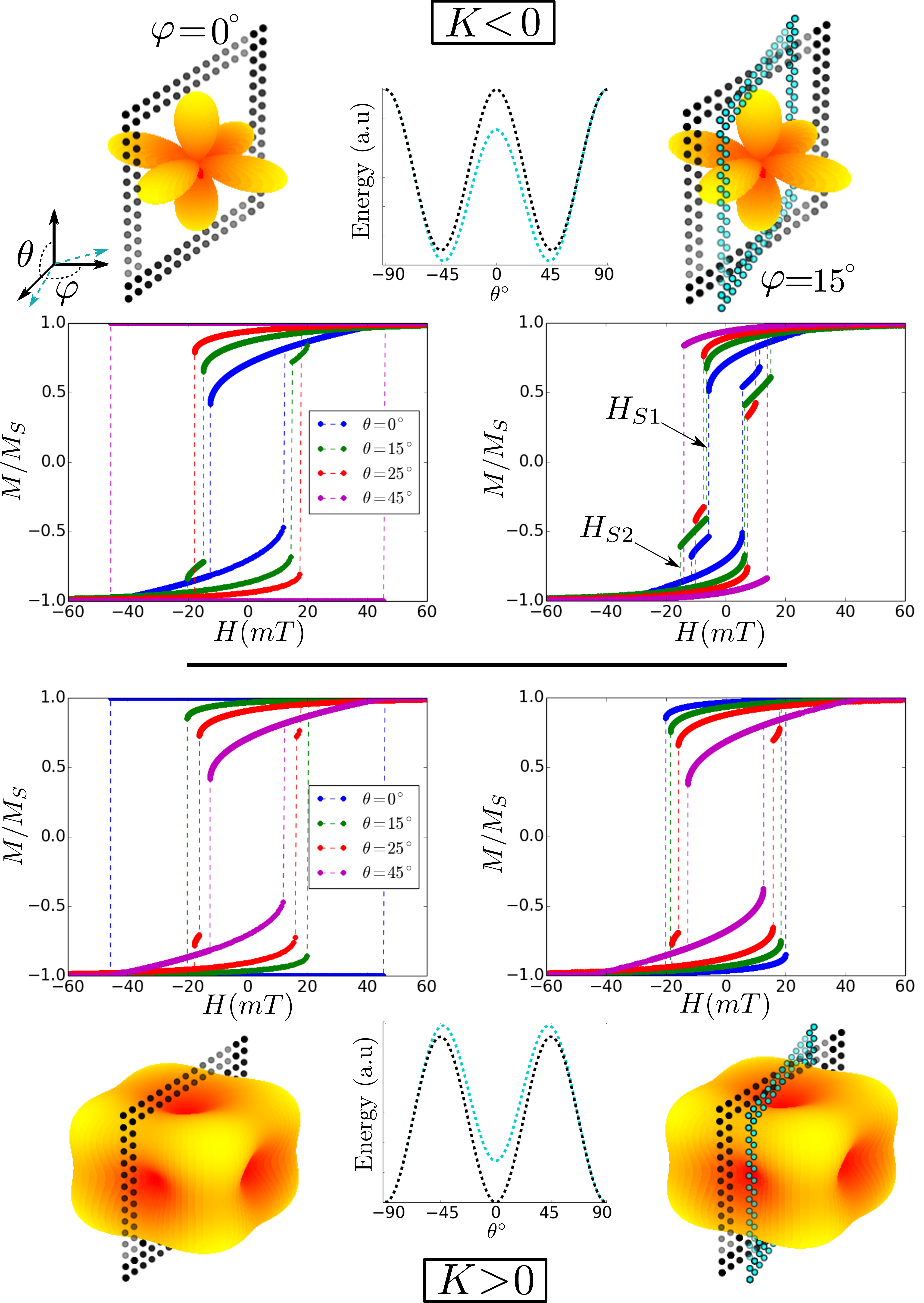}
\caption{Hysteresis loops corresponding to the extreme cases $x=0$ ($K_{C}<0$, top panel) and $x=1$ ($K_{C}>0$, bottom panel), for the ideal case $|K_{C}^{-}|=|K_{C}^{+}|$ (as described by equation \eqref{combined_NEG_POS-ideal}). For each case the loops correspond to rotation within the XZ plane ($\varphi=\ang{0}$; left panel) or at $\varphi=\ang{15}$ around the Z direction (right panel), to show the different symmetries throughout the loop. The magnetisation switching can be a one- or a two-step process (switching fields $H_{S1}$, $H_{S2}$).}
\label{fig:figure_2}
\end{figure}

Firstly, we have simulated the hysteresis loops of the reference cases corresponding to pure $K<0$ and $K>0$ (i.e. extreme cases $x=0$ and $x=1$). Typical hysteresis loops for various angles of field direction are shown in Figure \ref{fig:figure_2}, for two different rotation cross-sections: parallel to the XZ plane ($\varphi=\ang{0}$), and rotated ($\varphi=\ang{15}$) around the Z direction. 

\onecolumngrid

\twocolumngrid

The results displayed in Figure \ref{fig:figure_2} are similar -but of opposite trends- for both signs of anisotropy. Depending on the orientation between field and anisotropy easy axes, the magnetisation will undergo one or two switching events, and also exhibit different area and coercivity. For example, hysteresis loops for field variation along $\theta=\ang{0}$ have a strong dependence on the sign of the anisotropy. For $K>0$ the magnetisation starts in an easy direction giving a square loop and large coercivity, whereas for $K<0$ the magnetisation starts in a hard direction resulting in a rounded loop with reduced area and coercivity.

\begin{figure}[!ht]
\includegraphics[width = 1.0\columnwidth]{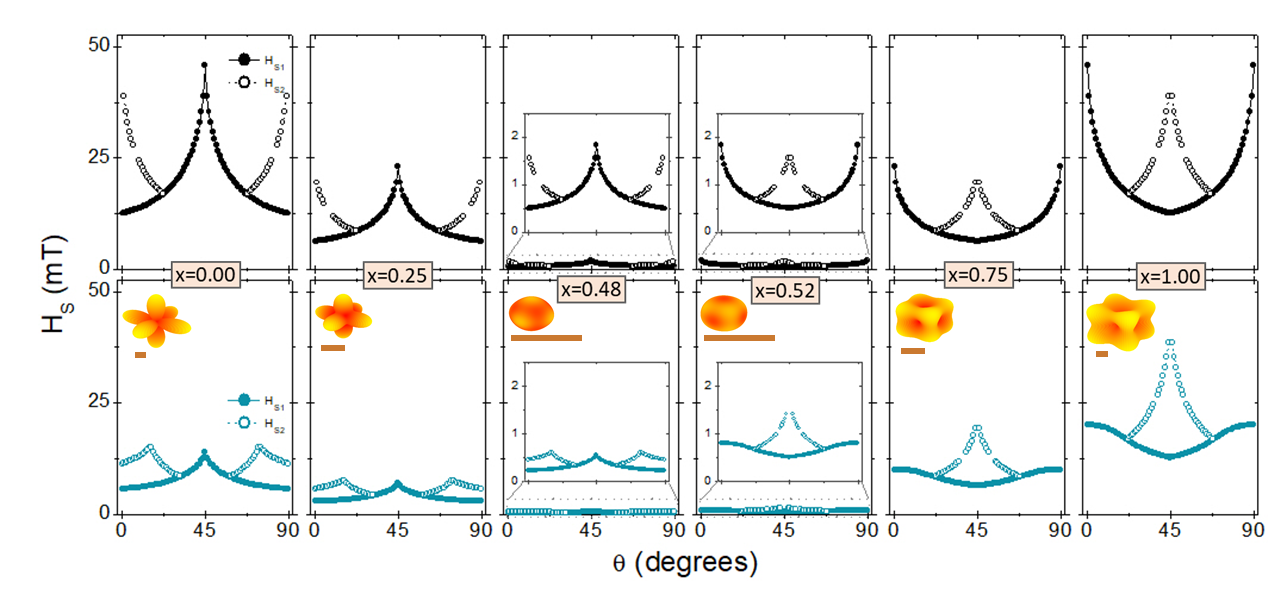}
\caption{Angular dependence of the switching fields (as described in Figure \ref{fig:figure_2}, for different fractions of positive and negative anisotropy constants. Top and bottom panels for each doping case correspond, respectively, to the $\varphi=\ang{0}$ and $\varphi=\ang{15}$ cases described in Figure \ref{fig:figure_2}. The yellow-orange drawings stand for the corresponding energy landscape of each $x$ fraction, with the close small bar illustrating the different energy scales.}
\label{fig:figure_3}
\end{figure}

Having shown and analyzed the known extreme cases of $x=0$ and $x=1$, we now proceed in a similar way (simulation of angular-dependent hysteresis loops) for mixed-anisotropy compositions as described by equation \eqref{combined_NEG_POS-ideal}. To compare the different cases we have focused on the angular-dependence of the switching fields ($H_{S}$), for the same illustrative cases of $\varphi=\ang{0}$ and $\varphi=\ang{15}$ of Figure \ref{fig:figure_2}. The results are summarized in Figure \ref{fig:figure_3}.

\onecolumngrid

\twocolumngrid

\begin{figure}[!h]
\includegraphics[width = 1.0\columnwidth]{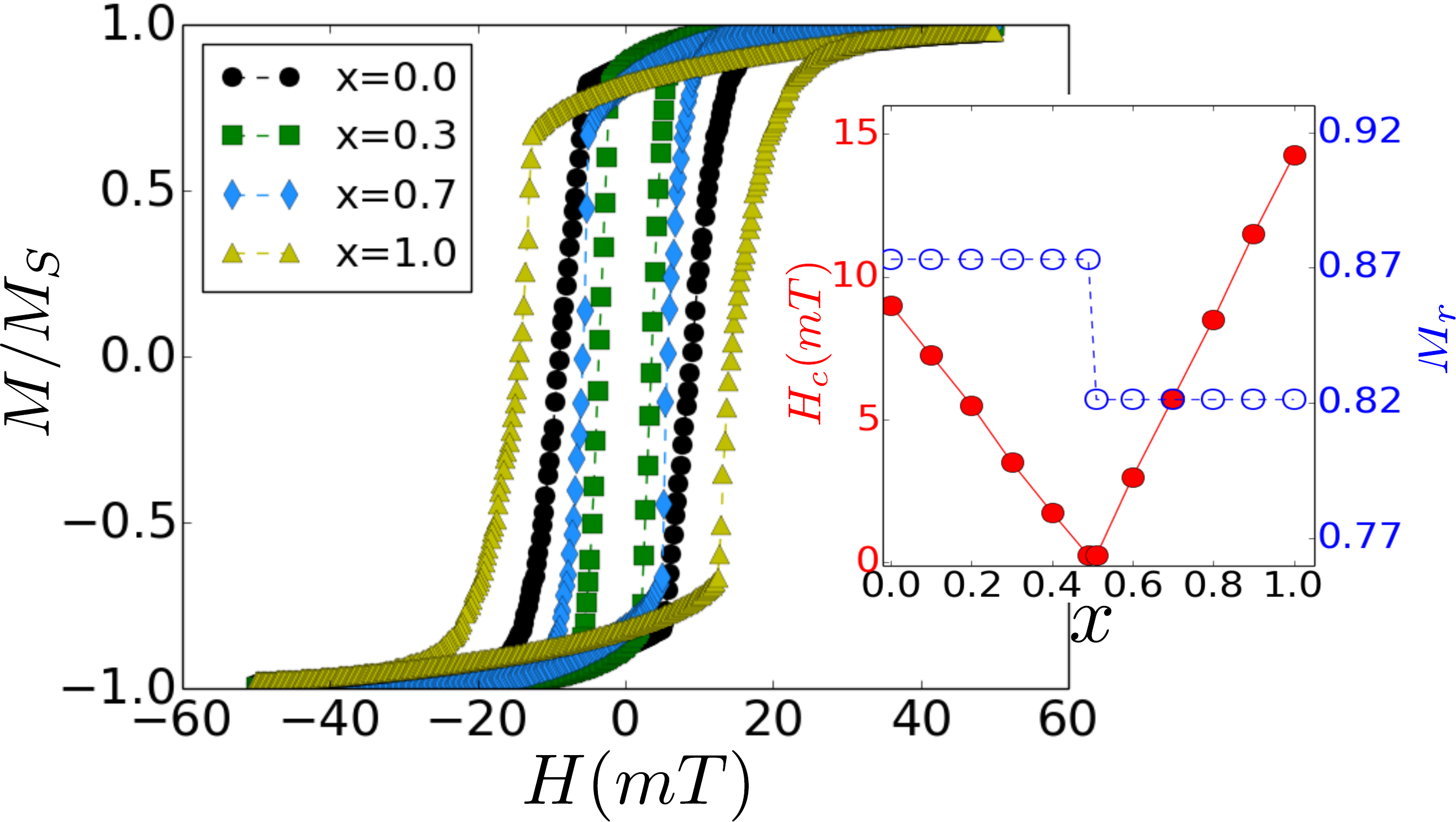}
\caption{Main panel: $M(H)$ hysteresis loops (corresponding to a randomly dispersed system), for various illustrative $Co_{x}Fe_{3-x}O_{4}$ compositions; inset: detailed evolution of the coercive field, $H_{c}$, and remanence, $M_{r}$, as a function of $x$.}
\label{fig:figure_4}
\end{figure}

The first main feature observed in Figure \ref{fig:figure_3} is that while the overall shape of the $H_{S}$ \textit{vs.} $\theta$ is quite similar for all $x$ cases, the absolute values rapidly decrease and tend to zero close to the $50\%$ composition (note that the choice of $x=0.48$ and $x=0.52$ was made specifically to illustrate this observation). 
The second main relevant aspect observed in Figure \ref{fig:figure_3} is that the apparent symmetry in the magnetic properties around the $x=0.50$ value observed for the case $\varphi=\ang{0}$ does not hold for the case of $\varphi=\ang{15}$. This is due to the different symmetry of the energy landscapes of the $K_{C}<0$ and $K_{C}>0$ cases (with 6 or 8 easy directions, respectively), which results in energy barriers of $K_{C}V/12$ or $K_{C}V/4$, respectively\cite{Yanes2007}. In order to weight the relative importance of such differences and considering the likely random orientation of a real sample in space, we have simulated the average properties of a system with randomly distributed easy axis directions for this ideal situation described by Eq. \eqref{combined_NEG_POS-ideal}. The results are displayed in Figure \ref{fig:figure_4}.

The main panel within Figure \ref{fig:figure_4} shows various hysteresis loops, including the theoretically well known extreme cases of pure $K_{C}<0$ ($x=0$) and $K_{C}>0$ ($x$=1). A detailed analysis of those indicates that the theoretical values of coercive field ($H_{c}$) and remanence ($M_{r}$) are well reproduced: for $K_{C}<0$ we obtain $H_{c}=0.20H_{K}$, where $H_{K}=2K/M_{S}$, and $M_{r}=0.87M_{S}$; and $K_{0}>0$ we find $H_{c}=0.32H_{K}$ and $M_{r}=0.82M_{S}$, as theoretically described by Usov \textit{et al.}\cite{Usov1997}. Such good agreement adds further support to our numerical results. 

Regarding the combination of anisotropies, two main characteristics are observed: on the one hand, $H_{C}$ shows a markedly linear initial decrease until reaching zero at about $x=0.5$, followed by a subsequent increase (also essentially linear, but of different slope); the extreme values in both cases correspond to the theoretical ones reported by Usov \textit{et al.}\cite{Usov1997} described earlier on. Since the coercive field is directly proportional to the anisotropy (no shape or interparticle interactions are present) this means that the approach predicts that \textit{the combination of opposite-symmetry anisotropies may lead to the cancellation of the effective total anisotropy}. We shall come back to this aspect later on. On the other hand and regarding the remanence, we also observed a two-regime behaviour but it this case of completely different features: $M_{R}$ is a bi-valued constant with the threshold at $x=0.5$, reaching the corresponding extreme cases for smaller and larger values as reported previously.

\subsection{The real $Co_{x}Fe_{3-x}O_{4}$ case: from $K_{C}^{-}=-1.1\cdot{10^4} J/m^3$ to $K_{C}^{+}=2.0\cdot{10^5} J/m^3$}\label{real_case}

\begin{figure}[h]
\includegraphics[width = 1.0\columnwidth]{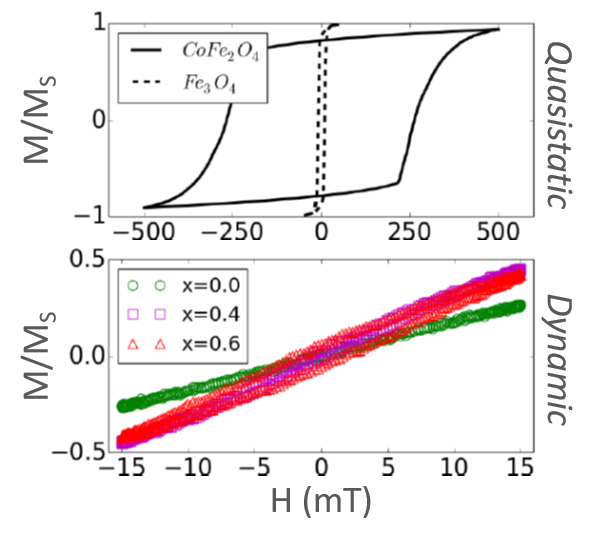}
\caption{Top panel: quasistatic $M(H)$ curves of the extreme cases of the series. Bottom panel: several dynamic $M(H)$ loops corresponding to the experimental conditions reported in Ref. \cite{fantechi2015} (frequency 183 kHz; field amplitude 12 kA/m; temperature 300 K).}
\label{fig:figure_5}
\end{figure}

Now that we have already analyzed the general properties demonstrated be the simple approach summarized by Eq. \eqref{combined_NEG_POS-ideal}, we investigate the behavior of the real $Co_{x}Fe_{3-x}O_{4}$ system as described by Eq. \eqref{combined_cubic_energies}. As mentioned in the Introduction, our goal is to study the hyperthermia performance of the $Co_{x}Fe_{3-x}O_{4}$ particles reported by Fantechi \textit{et al.}\cite{fantechi2015}. Thus, we simulated dynamic hysteresis loops for the experimental particle properties under the same field conditions, assigning to the effective anisotropy the value obtained from Eq. \eqref{combined_cubic_energies}. Some representative dynamical curves are shown in Figure \ref{fig:figure_5}, where also the quasistatic curves of the $x=0.0$ and $x=1.0$ limit cases are displayed for illustrative purposes.

\begin{figure}[h]
\includegraphics[width = 1.0\columnwidth]{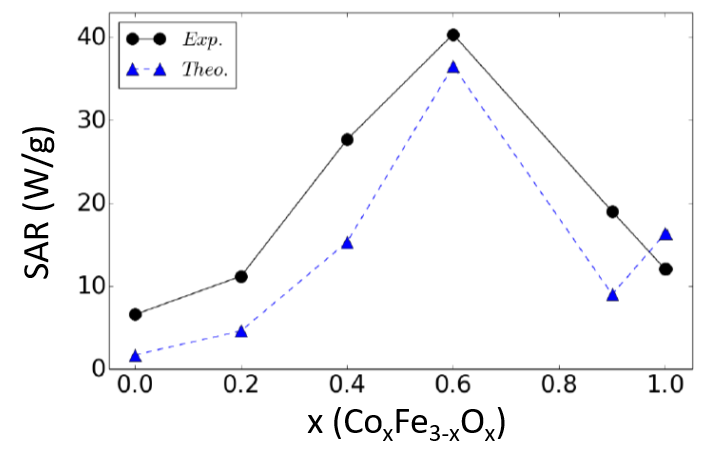}
\caption{Comparison between the experimentally measured $SAR$ values reported in Ref. \cite{fantechi2015} (black dots), and the ones obtained from the simulations (blue triangles).}
\label{fig:figure_6}
\end{figure}

The results displayed in Figure \ref{fig:figure_5} show that while a large difference depending on the composition is observed in the quasistatic curves (top panel), for the dynamical conditions (bottom panel) the curves become apparently very similar. However, the numerical evaluation of the hysteresis loop area of the dynamical loops indicate that there are in fact substantial differences. The loop area is used to estimate the \textit{Specific Absorption Rate} (SAR), i.e. the heating capability of the particles, usually defined as $SAR=area\times{frequency}$. The corresponding SAR values are summarized in Figure \ref{fig:figure_6}, showing good agreement between experiments and simulations. Such good agreement undoubtedly provides strong support to the simple effective anisotropy macrospin approach proposed here.

\begin{figure}[!ht]
\includegraphics[width = 1.0\columnwidth]{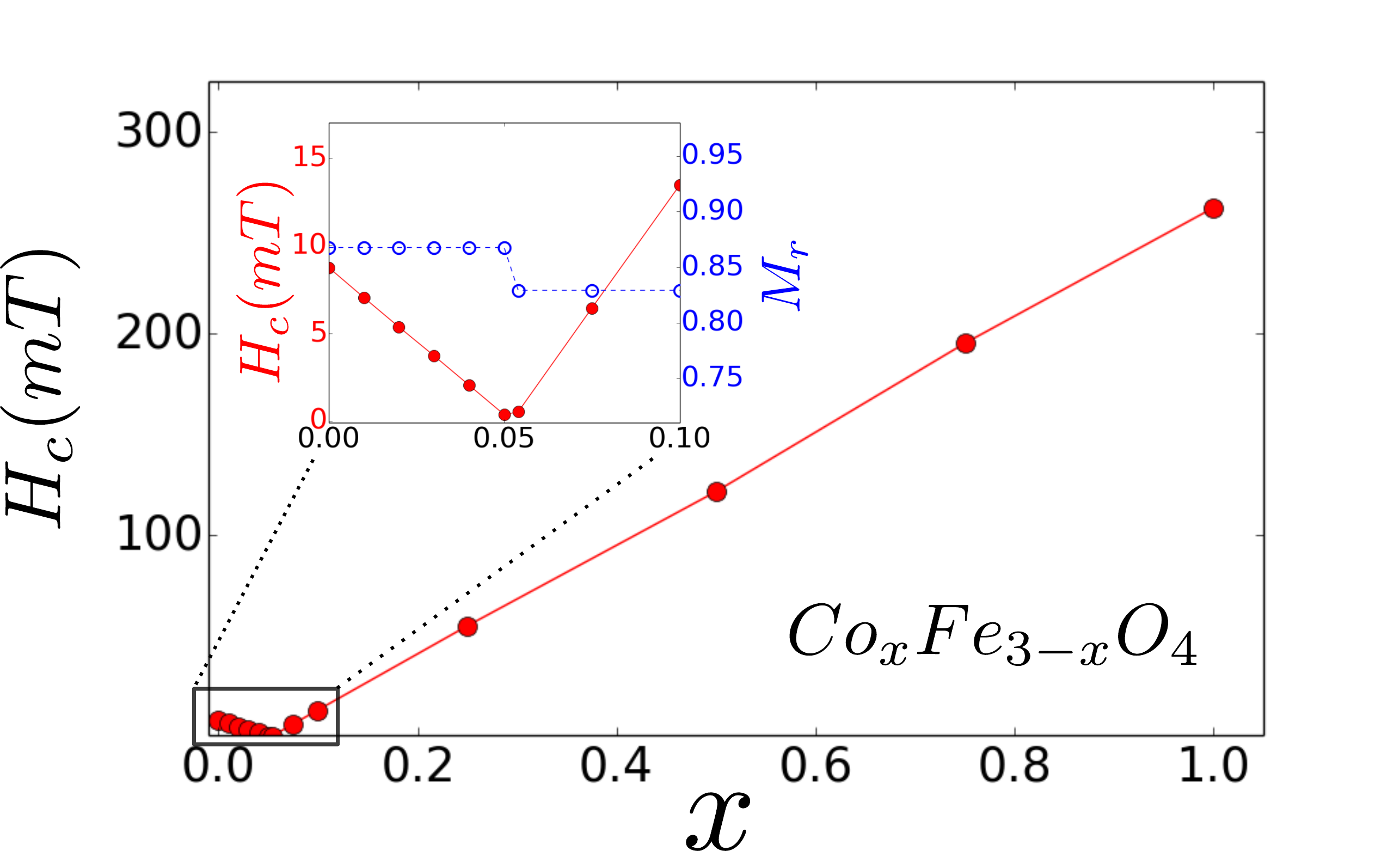}
\caption{Predicted $Co$-doping dependence of the coercive field for $Co_{x}Fe_{3-x}O_{4}$ nanoparticles according to Eq. \eqref{combined_cubic_energies}.}
\label{fig:figure_7}
\end{figure}

Having said that, the question posed in view of Figure \ref{fig:figure_4} still holds: would it be possible to lower the (already quite small)  anisotropy of $Fe_{3}O_{4}$ nanoparticles by doping with Co? This intriguing question is answered in Figure \ref{fig:figure_7}, which suggests that a small $Co$-doping (of about $5\%$) would lead to a suppression of the effective anisotropy.

\section{Conclusions}\label{conclusions}

We have considered theoretically, under the \textit{macrospin} approximation, the evolution of the effective magnetic anisotropy of the $Co_{x}Fe_{3-x}O_{4}$ series as a function of $x$, from the extreme cases of negative cubic-anisotropy of $Fe_{3}O_{4}$, to the positive cubic-anisotropy of $CoFe_{2}O_{4}$. Our phenomenological model assumes the $Co$-doping to result in an increasing $K_{C}>0$ contribution onto the $K_{C}<0$ $Fe_{3}O_{4}$ parent phase, which changes continuously both the magnitude and the symmetry of the net anisotropy. On this basis we have developed a simple model for the effective anisotropy of the macrospin in which   $K_{C}<0$ and $K_{C}>0$ coexist in combination,  directly proportional to the fraction of $Co$ present in the $Fe_{3}O_{4}$ phase. Applying this model to an experimental study on hyperthermia performance of $Co_{x}Fe_{3-x}O_{4}$ nanoparticles \cite{fantechi2015} we obtain a remarkably good agreement, which suggests that our simple assumptions may be reasonable. Interestingly, a side conclusion of our approach is that it predicts that a small $Co$-doping would lead to a decrease, and even the disappearance of the effective anisotropy, a quite unexpected result. It is the subject of future work to investigate such an intriguing possibility. However, we note that this may be complicated to demonstrate experimentally due to the required conditions, in particular it would be necessary to achieve highly spherical samples to remove shape-anisotropy effects, which for such small anisotropy could be very relevant and even dominate over the magnetocrystalline contribution\cite{Usov2010}). 

\section{Acknowledgements}
We acknowledge the Centro de Supercomputacion de Galicia (CESGA) for the computational resources. This research was partially supported by the  Conseller\'ia de Educaci\'on Program for Development of a Strategic Grouping in Materials (AeMAT)  at the Universidade de Santiago de Compostela (ED431E2018/08, Xunta de Galicia). D.S. also acknowledges Xunta de Galicia for financial support under the I2C Plan. Financial support of an International Exchanges grant (IE160535) of the Royal Society is gratefully acknowledged.

\bibliography{references}

\end{document}